\documentclass{ws-procs9x6_mod}

\begin{document}

\title{The ATLAS Transition Radiation Tracker}

\author{V.~A.~MITSOU}

\address{European Laboratory for Particle Physics (CERN), EP Division, \\
CH-1211 Geneva 23, Switzerland\\
E-mail: Vasiliki.Mitsou@cern.ch}

\address{\rm On behalf of the ATLAS TRT Collaboration}

\maketitle

\abstracts{ The Transition Radiation Tracker is a combined tracking and
electron identification detector, which is part of the ATLAS Inner Detector at
the CERN LHC. Tracking is carried out by drift tubes, while the interleaved
radiators produce detectable X-rays when electrons traverse them. The status of
the construction and quality control tests will be discussed. The results of
beam tests performed on detector prototypes and the expected performance in
ATLAS will also be reviewed. }

\section{Introduction}

The Transition Radiation Tracker (TRT) is a drift-tube system that, together
with silicon-strip and pixel detectors at low radii, constitutes the tracking
system of ATLAS, the Inner Detector\cite{idtdr}. It is meant to provide robust
tracking information with stand-alone pattern recognition capability in the LHC
environment, to enhance the momentum resolution by providing track measurement
points up to the solenoid radius and to provide a fast level-2 trigger. By
integrating the transition-radiation signature, the TRT also provides
stand-alone electron/pion separation (Fig.~\ref{fig:trt}, left).

\section{Detector design}

The TRT consists of 370,000 cylindrical drift tubes (straws). Each straw (4~mm
diameter, made of Kapton$^{\circledR}$ with a conductive coating) acts as a
cathode and is kept at high voltage of negative polarity. In the centre of the
straw there is a 30~$\mu$m diameter gold-plated tungsten sense wire. The layers
of straws are interleaved with the radiators (polypropylene foils or fibres).
The straws are filled with a gas mixture based on xenon (70\%) ---for good
X-ray absorption--- with the addition of CO$_2$ (27\%) and O$_2$ (3\%) to
increase the electron drift velocity and for photon-quenching. The Inner
Detector lies in a superconducting solenoid magnet generating a 2-T magnetic
field parallel to the beam axis over most of its volume.

\begin{figure}[ht]
\centerline{\epsfysize=1.37in\epsfbox{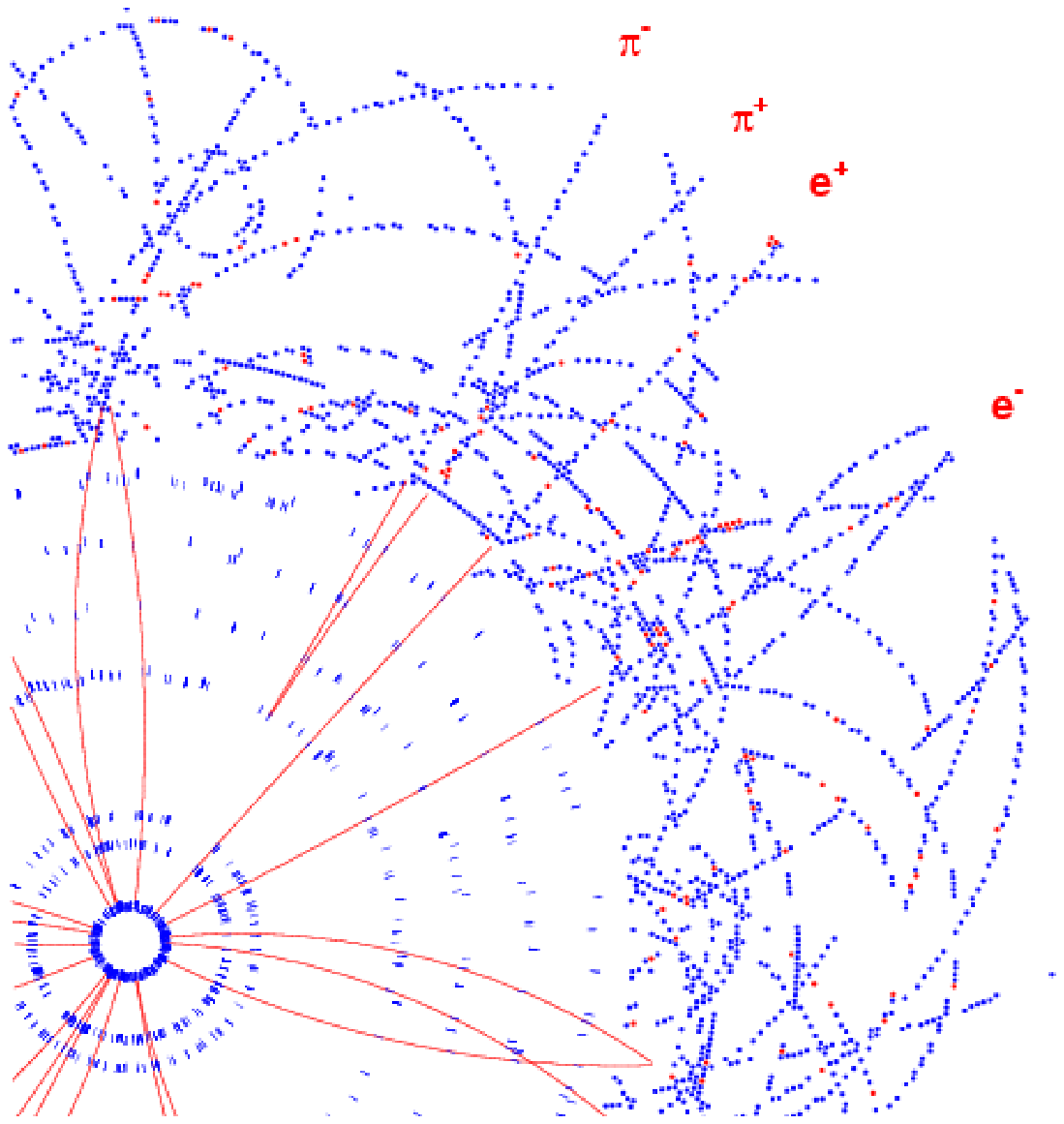}\hspace{1.5cm}\epsfysize=1.37in\epsfbox{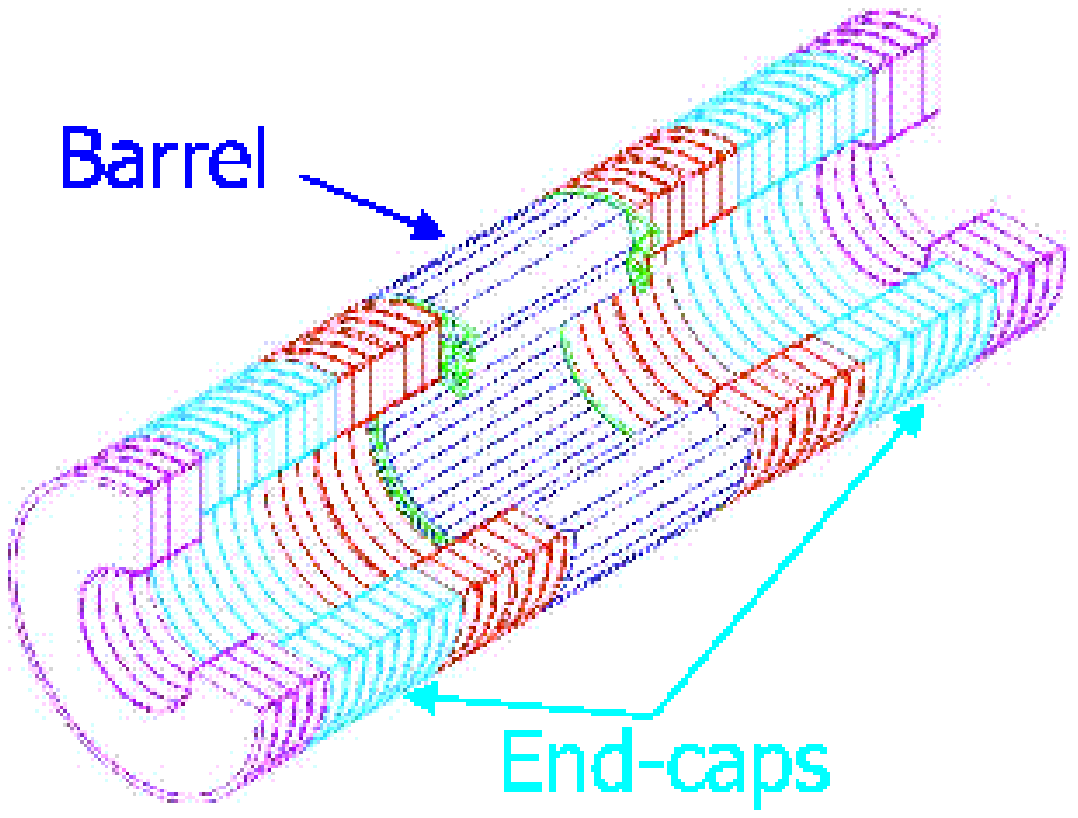}}
\caption{${\rm B_d^0\rightarrow J/\psi\;K_S^0}$ simulated event display in
the Inner Detector (left); schematic 3D-view of the TRT
(right).}\label{fig:trt}
\end{figure}

The TRT has three major modular components: the barrel and two end-caps
(Fig.~\ref{fig:trt}, right). The TRT barrel consists of three layers of 32
modules each. The straws are 144~cm long and are parallel to the beam occupying
the region between $56<r<107$~cm and $|z|<72$~cm, corresponding to a
pseudorapidity coverage of $|\eta|<0.7$\footnote{$r$ denotes the radius in the
plane transverse to the beam and $z$ the longitudinal position along the
beam.}. The wires are electrically split in the centre and are read out at both
ends, thus reducing the occupancy but doubling the number of electronic
channels.

In the end-caps, the straws are radially arranged in 18~units per side called
wheels, for a total of 224~layers of straws on each side ($83<|z|<340$~cm). The
straws extend between radii of 63~cm and 103~cm, except in the last four wheels
(64~layers), where the straws extend between radii of 47~cm and 100~cm,
covering thus a pseudorapidity range of $0.7<|\eta|<2.5$.

\section{Construction of the TRT}

The barrel module construction has taken place in the US (Duke, Hampton and
Indiana), whilst the construction of the end-cap wheels is performed in Russia
(PNPI and JINR). Assembled wheels and modules arrive at CERN, where they
currently undergo various acceptance tests before their integration and
installation into the ATLAS Inner Detector. The TRT construction is scheduled
to finish in 2005. The progress on the detector construction, the electronics
status and the TRT operation in a high-radiation environment are discussed in
some detail in Ref.~\refcite{Bari}.

\subsection{Barrel modules}

\begin{figure}[ht]
\centerline{\epsfysize=1.24in\epsfbox{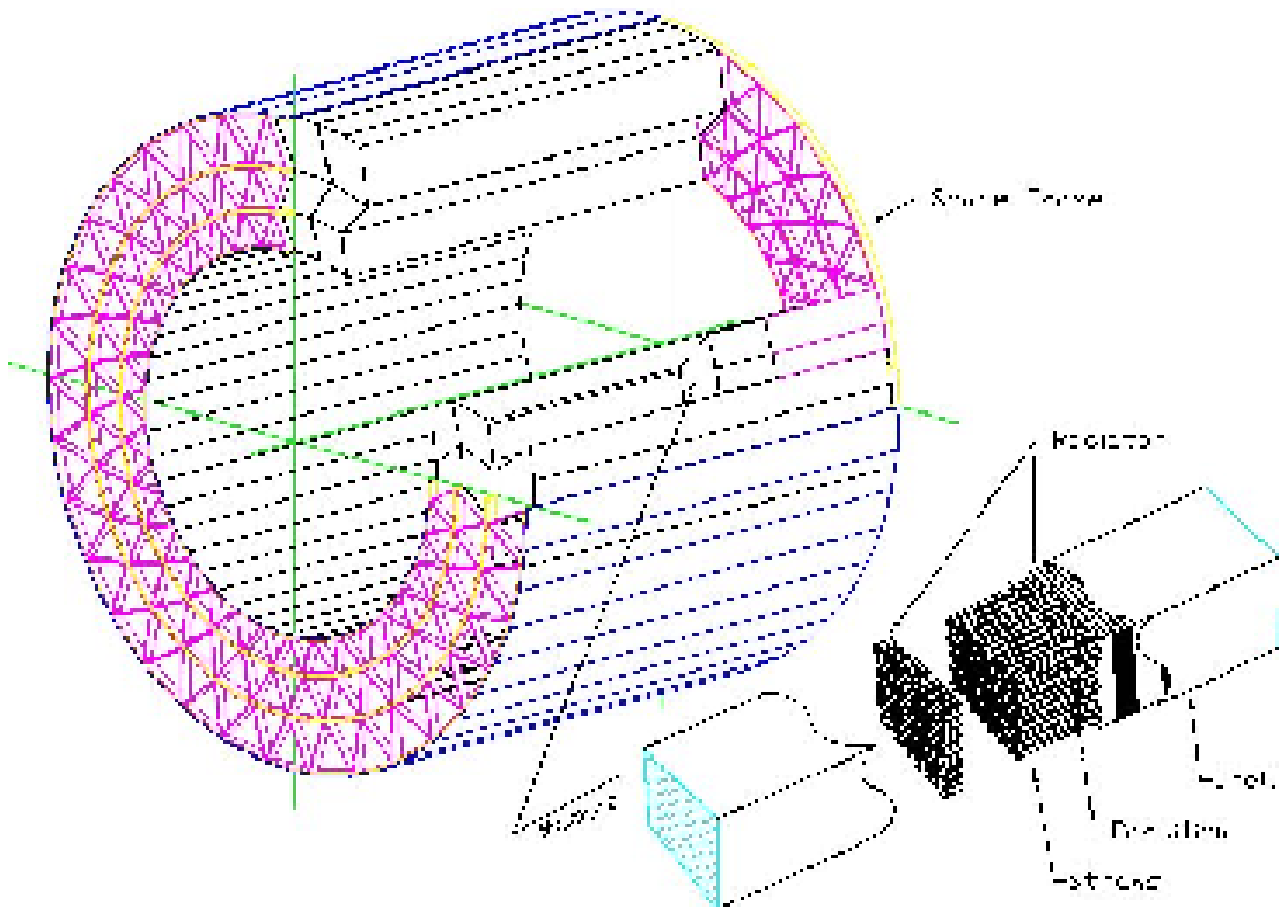}\hspace{1.3cm}\epsfysize=1.24in\epsfbox{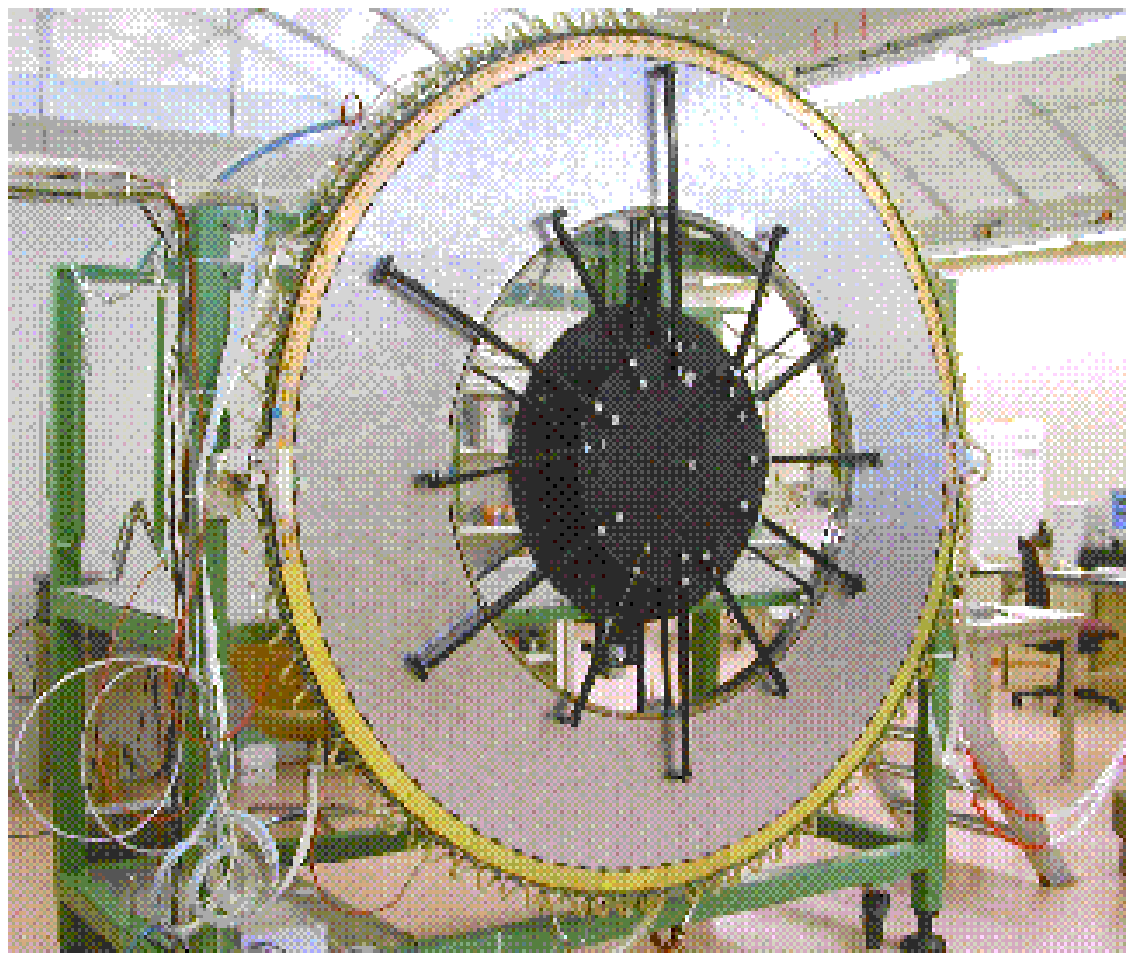}}
\caption{Schematic view of the barrel TRT and of a module (left); an
4-plane end-cap wheel on the X-ray scanner (right).}\label{fig:WTS}
\end{figure}

The assembly of the barrel modules (Fig.~\ref{fig:WTS}, left), including wire
stringing, has been completed over the last months and, for 40\% of them, the
gain mapping has been performed before shipping them to CERN. The gas gain is
measured at 50 points along each straw using X-rays\cite{x-ray}, to identify
wires that may be unstable for mechanical reasons (bent straws or misaligned
wires); these wires are then removed or replaced. The gain mapping, which lasts
2--3~days, is also used as a signal check and a long-term high voltage test by
recording the current draw.

The acceptance tests taking place at CERN include dimensional checks,
wire-tension measurements, gas-leak tests and high-voltage conditioning. Gain
mapping will also be performed together with possible repairs. In early 2004,
module assembly into the barrel support structure will begin and the front-end
electronics boards will be installed and commissioned.

\subsection{End-cap wheels}

Wheel production in Russia is making good progress despite some serious delays
with some of the components. Ten 4-plane wheels (out of a total of 80) have
been delivered to CERN and undergone acceptance tests, namely dimensional,
wire-tension, gas-tightness and high-voltage measurements as well as detailed
gain mapping of each straw. The first 8-plane wheel has been prepared and will
be used to begin the end-cap stacking procedure.

As mentioned above, the gas-gain uniformity at each straw is inspected in an
X-ray apparatus (Fig.~\ref{fig:WTS}, right). Through the gain scan, various
straw characteristics can be assessed, such as the straightness of the straws
or the gas-flow uniformity. The measurements, performed over $\sim$30,000
straws, have shown that more than 99.9\% of all straws are within the
specifications.

\section{Beam-test performance}

\subsection{Drift-time measurements}

\begin{figure}[ht]
\centerline{\epsfysize=1.14in\epsfbox{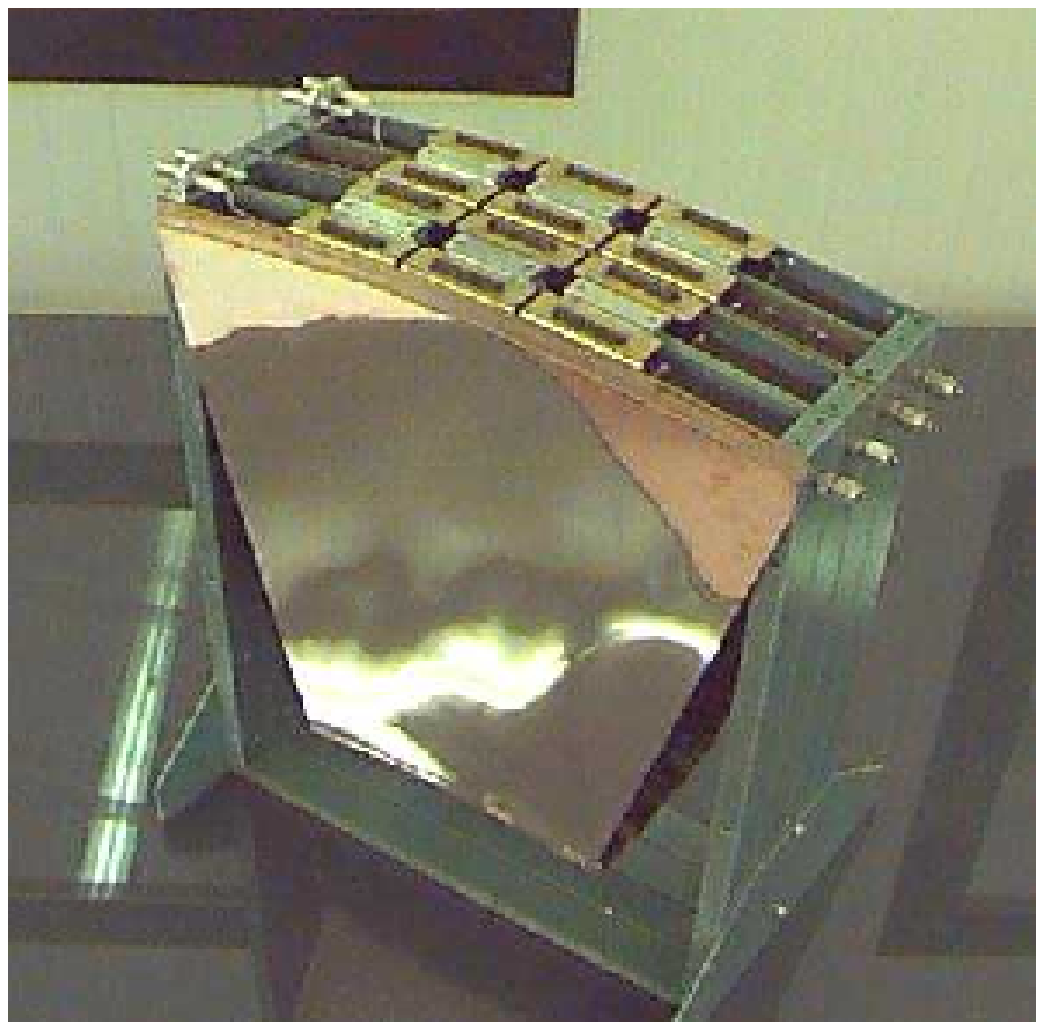}\hspace{1.1cm}\epsfxsize=2.0in\epsfysize=1.14in\epsfbox{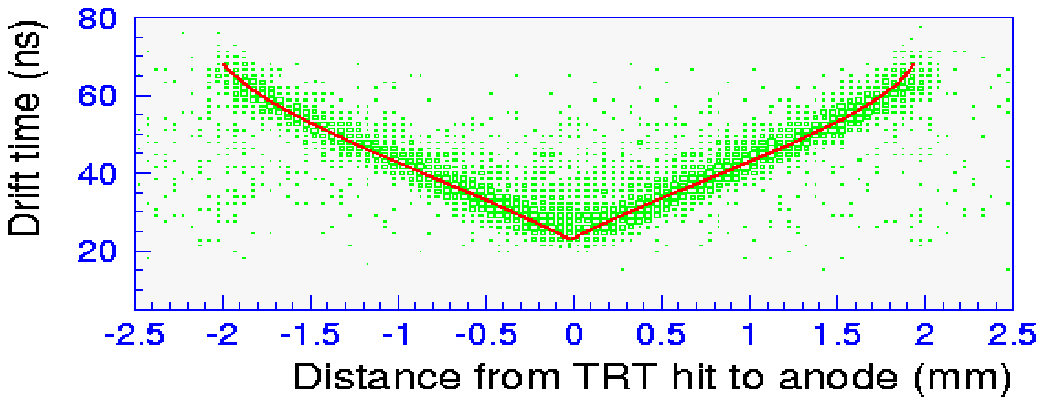}}
\caption{Wheel sector prototype with 16~layers of straws (left); measured drift
time versus distance of extrapolated beam track from the wire
(right).}\label{fig:drift}
\end{figure}

The position accuracy and efficiency\cite{drift-time} of the TRT have been
measured in a wheel sector equipped with prototype front-end boards
(Fig.~\ref{fig:drift}, left) exposed to a beam of high-energy particles at the
CERN SPS. The measured drift time is compared to the extrapolated position of
the beam track, determined with high accuracy using a Si-microstrip telescope.
The $r\!-\!t$ relationship in the straw is obtained in this way
(Fig.~\ref{fig:drift}, right) and is used to convert the drift-time measurement
into a radial distance from the sense wire. The parameters obtained from the
fit are fixed for a given gas gain, electronics signal shape and average energy
deposition in the straw, although they may vary somewhat from straw to straw
(straw eccentricity, non-circularity).

The maximum drift time in the gas is about 40~ns. For a 250-eV threshold, a
drift-time resolution of $\sim\!130~\mu$m has been achieved with an efficiency
of $\sim\!\!87\%$ (the basic straw hit efficiency is around 96\%)\cite{Bari}.

\subsection{Particle identification}

\begin{figure}[ht]
\centerline{\epsfysize=1.42in\epsfbox{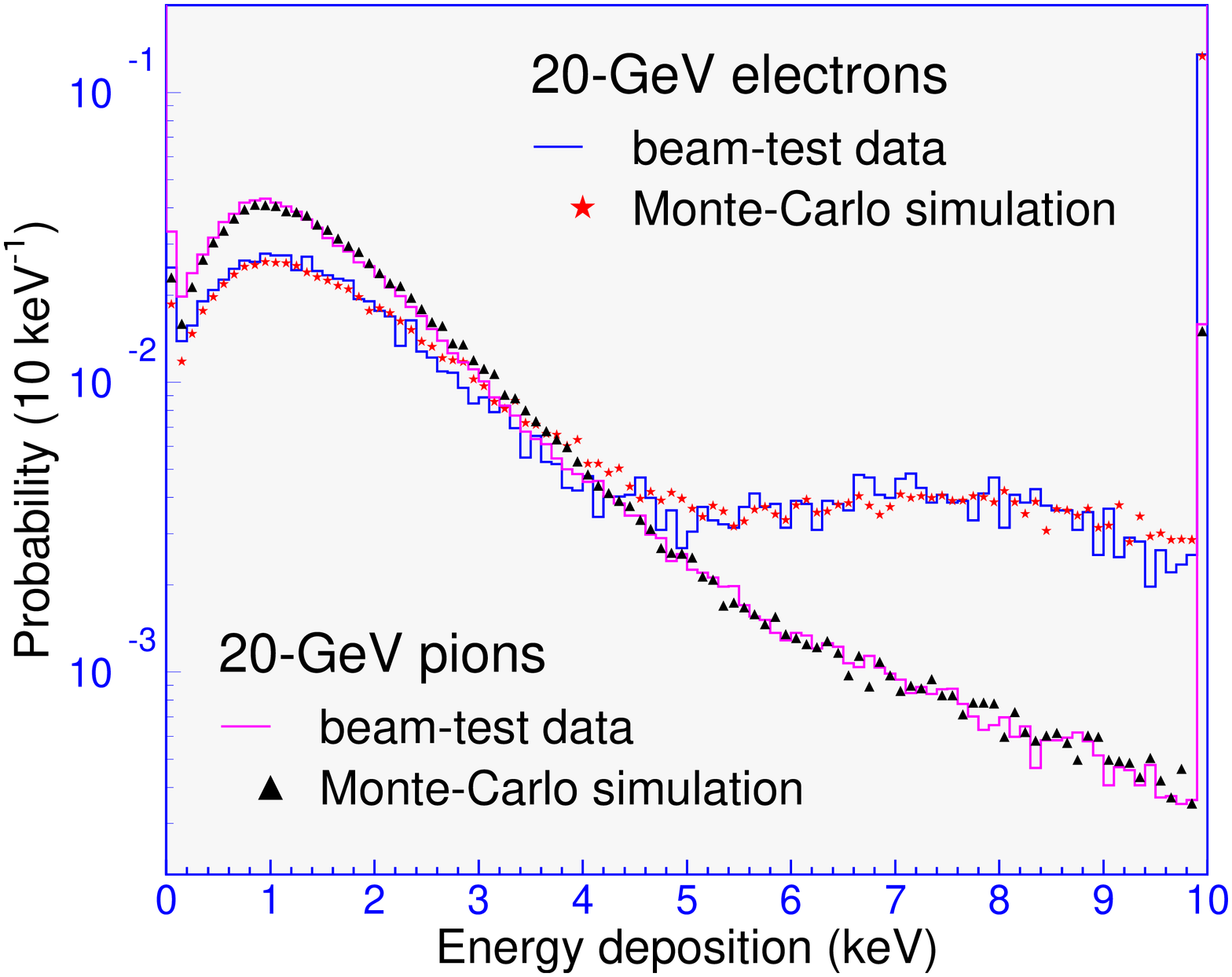}\hspace{1.1cm}\epsfysize=1.42in\epsfbox{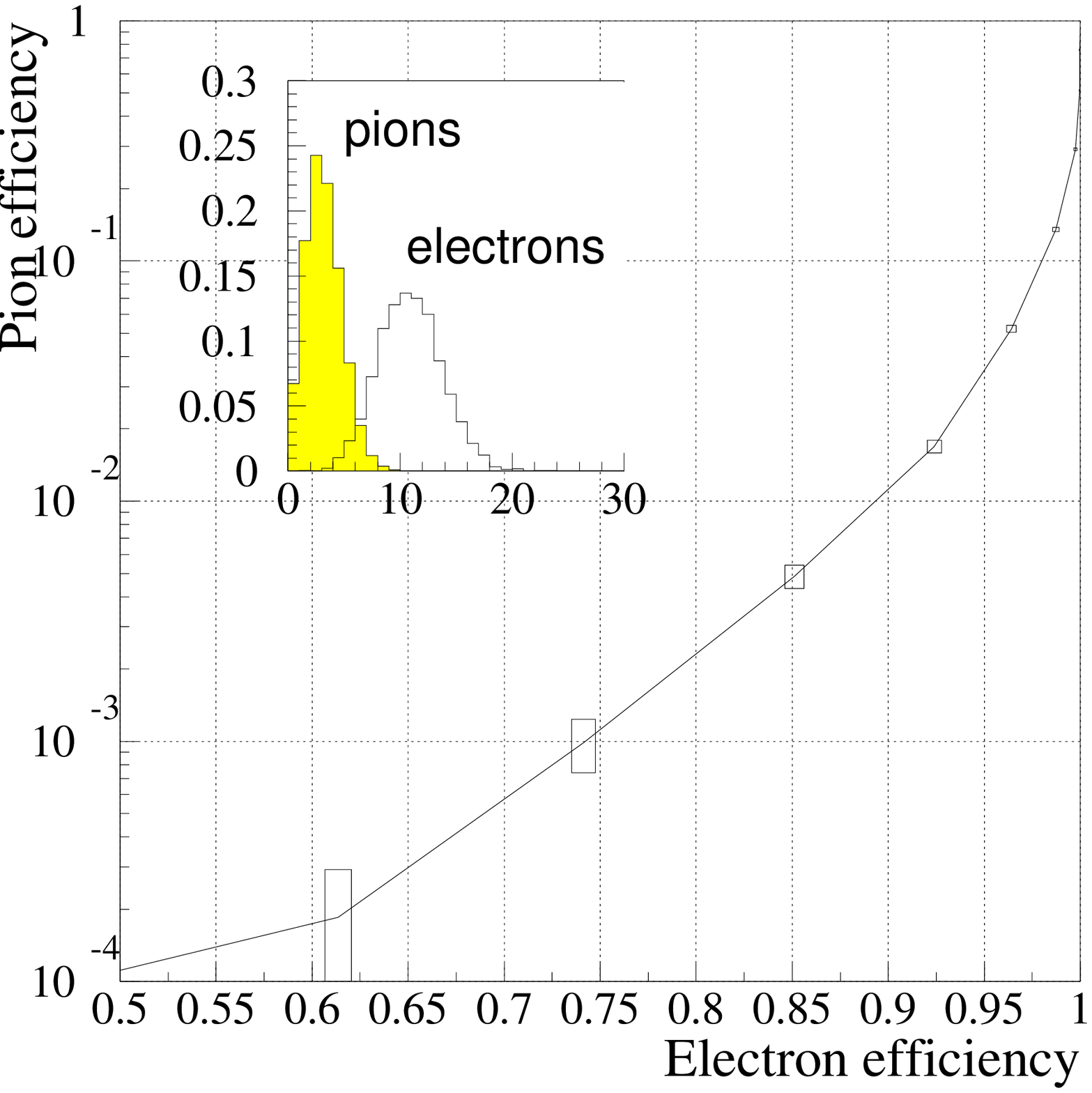}}
\caption{Differential energy spectra from data and simulation for a
 single straw with radiator (left); pion efficiency versus electron
 efficiency at 20~GeV (right).}\label{fig:pion_rej}
\end{figure}

Electron identification\cite{identification} makes use of the large energy
depositions due to transition radiation (TR). Typical TR photon energy
depositions in the TRT are 8--10~keV, while minimum-ionizing particles, such as
pions, deposit about 2~keV (Fig.~\ref{fig:pion_rej}, left). The parameter used
in electron identification is the number of local energy depositions on the
track above a given threshold.

The distributions of the number of energy depositions for pion and electron
tracks reconstructed using a wheel sector prototype are shown for a threshold
of $\sim\!\!6$~keV in the top left-hand corner of Fig.~\ref{fig:pion_rej}
(right). In the same figure, the resulting pion efficiency as a function of the
chosen electron efficiency is also displayed. For an electron efficiency of
90\%, the measured pion efficiency is about 1.2\%, i.e.\ a rejection factor of
75 is achieved against 20-GeV pions in a magnetic field of 0.8~T.

More recent beam-test results on the drift-time measurements and the
particle-identification performance of the TRT are discussed in
Ref.~\refcite{Bari}.

\section{Conclusions}

The construction of the ATLAS TRT is proceeding well. Significant progress on
the quality control of the main components and on the validation tests of the
modules and wheels has been made. The assembly of barrel modules is complete
and installation into the barrel support structure, to form the complete TRT
barrel, will begin in early 2004. The end-cap wheel production, on the other
hand, will only be completed in 2005. The quality-control procedures show that
the detector will operate initially with less than 1\% dead channels.

The beam-test results show that the TRT performance meets the requirements. A
drift-time resolution of $\sim\!130~\mu$m with an efficiency of 87\% is
feasible. For an electron efficiency of 90\%, the measured pion efficiency is
about 1.2\%, i.e.\ a rejection factor of 75 is obtained against 20-GeV pions.

\end{document}